\newcommand{\diracslash}[1]{#1\llap{/\kern2pt}}

\newcommand{\be}{\begin{equation}}
\newcommand{\ee}{\end{equation}}
\newcommand{\bea}{\begin{eqnarray}}
\newcommand{\eea}{\end{eqnarray}}
\newcommand{\ba}[1]{\begin{array}{#1}}
\newcommand{\ea}{\end{array}}

\documentclass[12pt]{iopart}
\usepackage{graphicx}
\begin{document}

\title{Pair-correlation in Bose-Einstein Condensate and
Fermi Superfluid of Atomic Gases}

\author{Bimalendu Deb}

\address{ Physical Research Laboratory, Navrangpura, Ahmedabad 380 009,
India}
\begin{abstract}
We describe pair-correlation inherent in the structure of
many-particle ground state of quantum gases, namely, Bose
Einstein condensate and Cooper-paired Fermi superfluid  of atomic
gases.  We make a comparative study on the pair-correlation
properties of these two systems. We discuss how to probe this
pair-correlation by stimulated light scattering. This intrinsic
pair-correlation may serve as a resource for many-particle
entanglement.
\end{abstract}


\def\be{\begin{equation}}
\def\ee{\end{equation}}
\def\bearr{\begin{eqnarray}}
\def\eearr{\end{eqnarray}}
\def\zbf#1{{\bf {#1}}}
\def\bfm#1{\mbox{\boldmath $#1$}}
\def\hf{\frac{1}{2}}

\pacs{03.75.Ss,74.20.-z,32.80.Lg}

\maketitle

\section{Introduction}

The realization of Bose-Einstein condensation in dilute atomic
gases \cite{bec} a decade ago marked a breakthrough revitalizing
many areas of physics, particularly, atomic and molecular
physics. One of the most significant advantages of
experimentation with cold atoms is the ability to tune  atom-atom
interaction over a wide range by a magnetic field Feshbach
resonance. This provides an unique opportunity to explore physics
of interacting many-particle systems in a new parameter regime.
In this context,
 cold atoms obeying
Fermi-Dirac statistics have currently attracted enormous research
interest. Fermions are the basic constituents of matter,
therefore research with trapped Fermi atoms
\cite{jin,hulet,thomas,ketterle} under controllable physical
conditions has important implications in materials science. In
particular, it has significant relevance in the field of
superconductivity.

The first achievement of quantum degeneracy in Fermi gas of
$^{40}$K atoms by Colorado group \cite{jin} in 1999 marked a
turning point in the research with cold atoms. Since then, cold
Fermi atoms have been of prime research interest in physics today.
In a series of experiments, several groups
\cite{hulet,thomas,solomon,mit,italy,grimm} have demonstrated
many new features of degenerate atomic Fermi gases. In a recent
experiment, Ketterle's group \cite{ketterle} has realized
quantized vortices as a signature of Fermi superfluidity in a
trapped atomic gas.  Two groups \cite{gap1,gap2} have
independently reported the measurement of pairing gap in Fermi
atoms. Collective oscillations \cite{duke,innsbruck} which are
indicative of  the occurrence of Fermi superfluidity
\cite{stringari}
 have been previously observed.
The crossover \cite{nozrink,randeria,crossover} between BCS state
of atoms and BEC  of molecules formed from Fermi atoms has become
a key issue of tremendous research interest. Several groups have
achieved BEC \cite{molecules} of molecules formed from Fermi
atoms. There have been several other experimental \cite{expt} and
theoretical investigations \cite{theory} on various aspects of
interacting Fermi atoms.

Both atomic BEC and superfluid atomic Fermi gas  have some common
quantum features: (a) both are macroscopic quantum objects (b)
the thermal de-Broglie wave-length greatly exceeds the
interparticle separation; (c) both have off-diagonal long range
order (ODLRO) or coherence; (d) the ground state of both the
systems has a structure whose constituents include pair-correlated
states; (e) both have ground state of broken symmetry; (f) both
must possess long wave-length phonon modes for restoration of
symmetry that is broken by their respective ground state. Our
focus here would be the common feature (d) to investigate how
this pair-correlation can be probed.

In the next section, we make a comparative study between BEC and
BCS ground states. Our objective is to show that a nontrivial
pair-correlation naturally arises in BEC \cite{deb4} and BCS
matter, and possibly it is a generic feature of all macroscopic
quantum objects.  In  section 3, we discuss briefly some relevant
features of trapped Fermi gas. In subsequent sections, we describe
stimulated light scattering as a means of probing
Cooper-pairing.  We find that using stimulated scattering of
circularly polarized light, it is possible to scatter selectively
either partner atom of a Cooper-pair \cite{deb3}. In the low
momentum transfer regime, this may be useful in exciting
Anderson-Bogoliubov phonon mode of broken symmetry.

\section{A comparison between BEC and BCS states}

Bose-Einstein condensate (BEC) of a weakly interacting Bose gas
and Bardeen-Cooper-Schrieffer (BCS) state of an interacting Fermi
gas are important in studies of macroscopic quantum physics. Both
refer to special states of matter in which conspicuous quantum
effects appear on a macroscopic scale. Both are quantum
degenerate matter. Quantum degeneracy refers to a physical
situation in which thermal de-Broglie wavelength of matter wave
exceeds inter particle separation. As a result, matter wave
properties play a crucial role in determining not only the
microscopic nature but also  the bulk properties of matter.
Particle-particle interaction in degenerate Bose and Fermi gas
leads respectively to Bose and Fermi superfluidity.

Let us  now  discuss some striking similarities as well as
differences in BEC and BCS ground states of interacting systems.
Let us begin by writing the ground states of uniform interacting
systems in momentum space
\begin{eqnarray} &\rm{BEC:}& \hspace{0.5cm} \Psi_0^{\rm{BEC}} =
\prod_{k\ne0} \phi_k=
\prod_{k\ne0}\frac{1}{u_{k}}\sum_{n=0}^{\infty}\left
(-\frac{v_{k}}{u_{k}}\right )^{n} \mid n_k,n_{-k} \rangle
\label{becgs} \end{eqnarray}
\begin{eqnarray} &\rm{BCS:}& \hspace{0.5cm} \Psi_0^{\rm{BCS}} =
\prod_{k}\psi_k = \prod_{k}\left (u_{k} \mid \mathbf{0} \rangle +
v_{k} \mid \mathbf{1}_{k\uparrow},\mathbf{1}_{-k\downarrow}
\rangle \right ) \label{bcsgs}
\end{eqnarray}
where $u_{k}$ and  $v_{k}$ are amplitude of corresponding Bose or
Fermi quasiparticle associated with celebrated transformation
that bears Bogoliubov's name. BEC ground state as expressed in Eq.
(\ref{becgs}) is a product of all possible nonzero momentum states
$ \phi_k $ which is a coherent superposition of two mutually
opposite momentum states $\mathbf{k}$ and $-\mathbf{k}$ occupied
by equal number of particles $n$ ranging from zero to infinity.
In other words, $\phi_k$ is a superposition of  all possible pair
states $\mid n_k, n_{-k} \rangle$. All the nonzero momentum
states compose the non-condensate part of BEC, while
zero-momentum state is the condensate part. Clearly, nonzero
momentum states form the structure in the ground state. At zero
temperature, non-condensate part consisting of nonzero momentum
states arises because of particle-particle interaction.
Therefore, we can infer that interaction leads to nontrivial
pairing correlation which may be used as a resource for generation
of continuous variable entanglement. How to extract this
correlation by light scattering and thereby to entangle two
spatially separated BECs in number and phase variables by a pair
of common laser beams passing through both the condensates has
been discussed elsewhere \cite{deb2}. Similar experimental
configuration has been recently used to produce and subsequently
measure phase difference between two spatially separated BECs
\cite{phase}.

\begin{figure}
 \includegraphics[width=4in,height=3.0in]{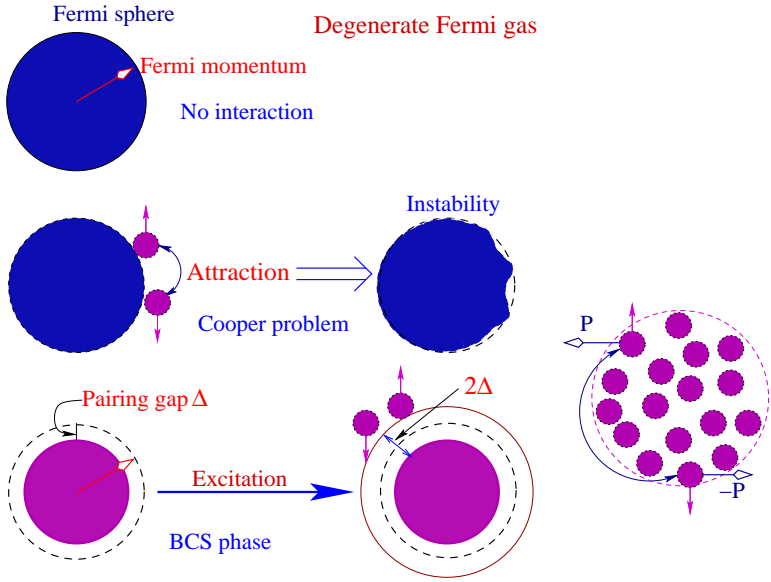}
 \caption{A naive pictorial illustration of Cooper-pair formation
 and its consequence. For a noninteracting (ideal) Fermi gas, the ground
 state is simply the Fermi sphere which is completely filled up to
 Fermi surface and completely empty above the surface. As
 shown first by Cooper, an attractive inter-fermion interaction,
 even if it is very weak, leads to formation of an exotic pair-bound state (Cooper-pair)
 which in turn leads to instability in the Fermi surface. Note that this
 pairing is basically a many-body effect, since for this effect to
 occur, quantum degeneracy or near degeneracy is essential.
 Bardeen, Cooper and Shrieffer then demonstrated that the ground
 state of a Fermi system with an attractive inter-particle
 interaction has a gap $\Delta$ which is known as pairing gap.
 Naively speaking, this ground state forms a sphere in momentum
 space with a radius which is less than Fermi energy  by an
 amount equal to $\Delta$. To break Coopr-pairs and thereby to
 excite single-particle excitations, a minimum of $2 \Delta$ energy
 is required to be imparted on the system. However, various
 collective modes among which Bogoliubov-Anderson mode is most
 significant one can be excited below the gap energy.
 The sphere at the
 extreme right is to be considered in real space and drawn to
 illustrate the fact that the two particles whose distance may exceed
 enormously the average inter-particle separation can form the
 pairing state. }
 \label{figcoop}
 \end{figure}

Now let us turn our attention to Eq. (\ref{bcsgs}) which
expresses the ground state of an attractively interacting
spin-half Fermi system. Figure 1 shows  pictorially and very
naively what happens to the ground state of noninteracting Fermi
system when attractive interaction is switched on. Like BEC ground
state, it has a structure that is based on particle-particle
pairing (Cooper-pairing) in mutually opposite momentum, albeit in
opposite spin up ($\uparrow$) and down ($\downarrow$) states. The
structure of BCS ground state differs from that of BEC because of
Pauli's exclusion principle which forbids more than one fermion
to occupy a single quantum state. Hence in a uniform Fermi
system, there is only one particle having momentum $\mathbf{k}$
and spin up, if it has to form pairing with another particle with
opposite momentum and down spin, it will find only one such
partner particle. Since pairing occurs in opposite momentum
states, the center-of-mass (COM) momentum of a Cooper-pair is
zero. Furthermore, the pairing state is  in spin-singlet  and
hence antisymmetric with respect to spin degrees of freedom.
Therefore its spatial part must be symmetric. This means pairing
must occur in even number of relative angular momentum $l$. In
low temperature weak-coupling superconductor, Cooper-pairing
occurs in s-wave ($l=0$) state. Although, a Cooper-pair is a kind
of two-particle bound state, it is fundamentally different from
familiar bound states like diatomic molecule. Cooper-pairing is
basically a many-body phenomenon. It occurs only when fermions
attract one another under quantum degenerate condition. In
contrast a diatomic molecule can be formed by three body
interaction. A single molecule can exist in isolation. In
contrast, any attempt to isolate a single Cooper-pair from
many-body degenerate environment will result in its breaking up
into individual fermions. When molecule formation takes place,
only nearest neighbor particles form molecular bonding.
Cooper-paring can occur between two fermions lying far apart,
their distance can greatly exceed average inter-fermion
separation. Cooper-pairs  can condense into zero (COM) momentum.
In fact, a crossover from BCS state of atoms to BEC state of
molecules formed from atoms due to a magnetic field Feshbach
resonance is an important object of current research interest.

\section{BCS state of trapped Fermi gas of atoms}

To illustrate the main idea, we specifically consider trapped
$^6$Li Fermi atoms in their two lowest hyperfine spin states
$\mid g_1 \rangle = \mid 2{\rm S}_{1/2}, F = 1/2, m_{F} = 1/2
\rangle$ and $\mid g_2 \rangle = \mid 2{\rm S}_{1/2}, F=1/2, m_F
= -1/2 \rangle $. For s-wave pairing to occur, the atom number
difference $\delta N $ of the two components should be restricted
by $\frac{\delta N}{N} \le T_c/\epsilon_F$ where $T_c$ is the
critical temperature for superfluid transition and $\epsilon_F$
is the Fermi energy at the trap center.  Unequal densities of the
two components result in interior gap (IG) superfluidity
\cite{wilczek,mishra}. We have suggested in Ref. \cite{mishra}
that it is possible to  experimentally  realize IG state in
two-component Fermi gas of $^6$Li atoms by making density
mistmatch between the two spin-components. In two remarkable
recent experiments \cite{igexp1,igexp2} using two-component $^6$Li
gas, some results which indicate the occurrence of IG state have
been obtained.  We here consider only the case $N_{1/2} =
N_{-1/2}$ which is the optimum condition for s-wave Cooper
pairing.

Let us consider a cylindrical harmonic  trap characterized by the
radial (axial) length scale $a_{\perp (z)}=
\sqrt{\hbar/(m\omega_{\perp (z)}}$. One can define  a geometric
mean frequency
 $\omega_{ho} =
(\omega_{\perp}^2\omega_z)^{1/3}$ and a mean length scale by
$a_{ho} = \sqrt{\hbar/(m\omega_{ho})}$. In Thomas-Fermi local
density approximation (LDA) \cite{houbiers},  the state of the
system is governed by $ \epsilon_F({\mathbf r})+ V_{ho}({\mathbf
r}) + U({\mathbf r}) = \mu, $ where $\epsilon_F ({\mathbf r}) =
\hbar^2 k_F({\mathbf r})^2 /(2m)$ is the local Fermi energy,
$k_F({\mathbf r})$ denotes the local Fermi momentum which is
related to the local number density by $n({\mathbf r}) =
k_F({\mathbf r})^3/(6\pi^2)$. Here $U$ represents the mean-field
interaction energy and $\mu$ is the chemical potential. At low
energy, the mean-field interaction energy depends on the two-body
s-wave scattering amplitude $f_0(k)= -a_s/(1+ia_sk)$, where $a_s$
represents  s-wave scattering length and $k$ denotes the relative
wave number of two colliding particles. In the dilute gas limit
($|a_s|k <\!<1$), $U$ becomes proportional to $a_s$ in the form
$U({\mathbf r}) = \frac{4\pi\hbar^2 a_s}{2m} n({\mathbf r})$.
 In the unitarity limit $|a_s|k \rightarrow \infty$,
the scattering amplitude $f_0 \sim i/k$ and hence  $U$ becomes
independent of $a_s$. It then follows from a simple dimensional
analysis that in this limit, $U$ should be proportional to the
Fermi energy: $U({\mathbf r}) = \beta \epsilon_F({\mathbf r})$
where $\beta$ is the constant.  In this limit, the pairing gap
also becomes proportional to the Fermi energy.

Under LDA,  the density profile of a trapped Fermi  gas is given
by \bearr n({\mathbf r}) = n({\mathbf 0}) (1 -
r_{\perp}^2/R_{\perp}^2 - r_{z}^2/R_z^2 )^{3/2}, \label{nr} \eearr
where $ n({\mathbf 0}) = 1/(6\pi^2\hbar^3)[2m\mu/(1+\beta)]^{3/2}$
is the density of the atoms at the trap center. Here $R_{\perp
(z)}^2 = 2\mu/(m\omega_{\perp (z)}^2)$ is the radial(axial)
Thomas-Fermi radius. The normalization condition on eq. (\ref{nr})
gives an expression for  $
 \mu = (1+\beta)^{1/2} (6N_{\sigma})^{1/3}\hbar\omega_0
$ where $N_{\sigma}$  is the total number of atoms in the
hyperfine spin $\sigma$. The Fermi momentum $k_F = [3\pi^2
n({\mathbf 0})]^{1/3} = (1+\beta)^{-1/4}k_F^0$ where $ k_F^0 =
(48N_{\sigma})^{1/6}/a_{ho} $ is the Fermi momentum of the
noninteracting trapped gas.

\section{Stimulated light scattering in Cooper-paired Fermi atoms}

To unravel the nature of Fermi superfluid of atomic gases, it is
important to analyze the possible response of this quantum gas due
to an external perturbation.
 A method has been suggested to use
resonant light \cite{zoller} to excite one of the spin components
into an excited electronic state and thereby  making an interface
between normal and superfluid atoms as in superconductive
tunneling.  This  has a threshold equal to the gap energy
$\Delta$. This method has been applied in recent experiments
\cite{gap1,theogap1} with the use of rf field for  estimating gap
energy. There have been a number of proposals \cite{zoller,huletp}
for probing pairing gap.

We calculate response function of superfluid Fermi gas due to
stimulated light scattering that does not cause any electronic
excitation in the atoms. We particularly emphasize the role of
light polarization in single-particle excitations which have a
threshold $2\Delta$. We present a scheme by which it is possible
to have single-particle excitation in only one partner atom (of a
particular hyperfine spin state) of a Cooper-pair using proper
light polarizations in the presence of a magnetic field. This may
lead to better precision in spin-selective time-of-flight
detection of scattered atoms. Furthermore, spin-selective light
scattering allows for unequal energy and momentum transfer into
the two partner atoms of a Cooper-pair. This may be useful in
exciting Bogoliubov-Anderson (BA) phonon mode of symmetry breaking
by making small difference in momentum transfers received by the
two partner atoms from the photon fields. Recently, a number of
authors \cite{mottelson,griffin,minguzzi} have studied
Bogoliubov-Anderson (BA)  mode \cite{bamode,anderson,martin} in
fermionic atoms as a signature of superfluidity.  BA mode
 is associated with long wave  Cooper-pair
density fluctuations. In electronic superconductor, this mode is
suppressed due to long wave Coulomb interaction. In neutral
superfluid Fermi system such as trapped atomic Fermi gas, this
mode is well defined and should in principle be experimentally
observable. However, its experimental detection  poses a
challenging problem since it is a near zero-energy zero-momentum
mode.

Figure 2 shows the schematic level diagram for stimulated light
scattering by two-component $^6$Li atoms in the presence of an
applied magnetic field which is tuned near the Feshbach resonance
($\sim 834$ Gauss) results in strong inter-component s-wave
interaction. At such high magnetic fields, the splitting between
the two ground hyperfine states is $\sim 75$ MHz
 while the corresponding splitting between the
excited states $ \mid e_1 \rangle = \mid 2{\rm P}_{3/2}, F=3/2,
m_{F} = -1/2 \rangle $ and $\mid e_2\rangle = \mid 2{\rm
P}_{3/2}, F = 3/2, m_{F} = -3/2 \rangle $ is $\sim 994$ MHz. Two
off-resonant laser beams with a small frequency difference are
impinged on atoms, the scattering of one laser photon is
stimulated by the other photon. In this process, one laser photon
is annihilated and reappeared as a scattered photon propagating
along  the other laser beam. The magnitude of momentum transfer
is $q \simeq 2 k_L \sin(\theta/2) $, where $\theta$ is the angle
between the two beams and $k_L$ is the momentum of a laser
photon. Let  both the laser beams be $\sigma_{-}$ polarized and
tuned near the transition $\mid g_2\rangle \rightarrow \mid
e_2\rangle$. Then the transition between the states $\mid g_1
\rangle $ and $\mid e_2 \rangle $ would be forbidden while the
transition $\mid g_1 \rangle \rightarrow \mid e_1\rangle $ will
be suppressed due to the large detuning $\sim 900$ MHz. This
leads to a situation where the Bragg-scattered atoms remain in
the same initial internal state $\mid e_2\rangle$. Similarly,
atoms in state $\mid g_1 \rangle$ only would undergo Bragg
scattering when two $ \sigma_{+}$ polarized lasers are tuned near
the transition $\mid g_1 \rangle \rightarrow \mid 2{\rm P}_{3/2},
F = 3/2, m_F = 3/2\rangle$. Thus, we infer that in the presence
of a high magnetic field,  it is possible to scatter atoms
selectively of either spin components only by using circularly
polarized Bragg lasers. We assume that both the laser beams are
$\sigma_-$ polarized and tuned  near the transition $\mid
g_2\rangle \rightarrow \mid e_2 \rangle $. Under such conditions,
considering a uniform gas of atoms, the effective laser-atom
interaction Hamiltonian \cite{deb3} in electric-dipole
approximation is
 $H_I \propto  \sum_{{\mathbf k},\sigma=1,2}\gamma_{\sigma\sigma}
  \hat{c}_{\sigma}^{\dagger}({\mathbf k }+ {\mathbf q})
  \hat{c}_{\sigma}({\mathbf
  k})$,
where $\hat{c}_{\sigma}(k)$ represents annihilation operator of an
atom with momentum $\mathbf{k}$ in the internal state $\sigma$.
The subscript $\sigma=1(2)$ refers to the state $\mid g_1\rangle$
($\mid g_2\rangle$). The bare vertex $\gamma_{\sigma\sigma}$ is
given by the Kramers-Heisenberg formula \cite{sakurai} \bearr
\gamma_{\sigma\sigma} =
\frac{e^2\mathcal{E}_1\mathcal{E}_2}{m_e\sqrt{n}\hbar\omega_1\omega_2}
 \sum_{i=1,2}
\frac{(\mathbf{d}_{\sigma\sigma}.\hat{\mathcal{E}}_2)(\mathbf{d}_{\sigma\sigma}.\hat{\mathcal{E}}_1)}
{\hbar^2(\omega_{\sigma\sigma}- \omega_i)} \eearr where $d_{ii}$
is the dipole matrix element between  $\mid g\rangle_i$ and  $\mid
e\rangle_i$ and $n$ is the incident photon number which is assumed
to be equal for both the laser beams. Here $m_e$ and $e$ are the
mass and charge, respectively, of the valence electron;
$\hat{\mathcal{E}}_i$ and $\omega_i$ represent  the electric field
and frequency, respectively, of $i$-th laser beam and
$\omega_{\sigma\sigma}$ is the atomic frequency between the states
$\mid g\rangle_{\sigma}$ and $\mid e\rangle_{\sigma}$. For the
particular case of $\sigma_{-}$ polarization in the presence of
magnetic field as discussed above, one finds $\gamma_{22}>\!>
\gamma_{11}$.
 On the other hand, in the absence of magnetic field,  one has
$\gamma_{11} \simeq \gamma_{22}$.

\subsection{The response function}

 We assume that, except the
center-of-mass momentum, the spin or any other internal degrees of
atom does not change due to light scattering.

Now, one can define the density operators  by $\rho_q^{(0)} =
\sum_{\sigma,\mathbf{k}}
a_{\sigma,\mathbf{k}+\mathbf{q}}^{\dagger}a_{\sigma,\mathbf{k}}$
and \bearr
 \rho_q^{(\gamma)} = \sum_{k,\sigma}
 \gamma_{\sigma\sigma} a_{\sigma,\mathbf{k}+\mathbf{q}}^{\dagger}a_{\sigma,\mathbf{k}}
 \eearr
One can identify the operator $\rho_{q}^{(0)}$ as the Fourier
transform of the density operator in real space. The scattering
probability is related to the susceptibility \bearr
\chi(\mathbf{q},\tau-\tau') = -\langle
T_{\tau}[\rho_q^{(\gamma)}(\tau)\rho_{-q}^{(\gamma)}(\tau')]\rangle.
\eearr where $T_{\tau}$ is the complex time $\tau$ ordering
operator and $\langle \cdots \rangle$ means thermal averaging.
The dynamic structure factor is related to $\chi$ by
$\chi(\mathbf{q},\omega_n)$ as \bearr S(\mathbf{q},\omega) = -
\frac{1}{\pi}[1+n_B(\omega)]\rm{Im}[ \chi(\mathbf{q},z=\omega +
i0^{+})]. \label{flucdiss}\eearr This follows from generalized
fluctuation-dissipation theorem. In order to treat collective
excitations, it is essential to go beyond Hartree approximation
and apply either a kinetic equation or a time-dependent
Hartree-Fock equation  or a random phase approximation
\cite{martin}. The essential idea is to take into account the
residual terms which are neglected in the BCS approximation and
thereby treat the off-diagonal matrix elements (vertex functions)
of single-particle operators in a more accurate way
\cite{schrieffer,martin}.

\begin{figure}
 \includegraphics{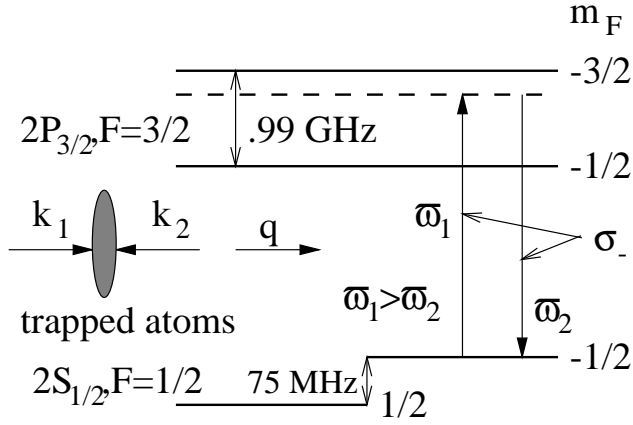}
 \caption{A schematic level diagram for polarization-selective light scattering
 in two-component Fermi gas of $^6$Li atoms}
 \label{diapol}
 \end{figure}

The detailed method of calculation of response function of
superfluid Fermi atoms due to stimulated  light scattering is
given elsewhere \cite{deb3}.

  We here present the final result
 \bearr \chi(\mathbf{q},\omega) =
2N(0)\gamma_0^2\langle B \rangle + 2 N(0)   \left [\langle A
\rangle + \frac{\omega^2 \langle f \rangle ^2 }{4\Delta^2\langle
\beta^2f\rangle}\right ] \gamma_3^2 \label{chinew}\eearr where
$N(0)$ is the density of states at the Fermi surface and
 \bearr A =
\frac{(\mathbf{v}_k.\mathbf{p}_q)^2-\omega^2
f}{\omega^2-(\mathbf{v}_k.\mathbf{p}_q)^2}, \hspace{0.5cm}
 B =
\frac{(\mathbf{v}_k.\mathbf{p}_q)^2(1-f)}{\omega^2-(\mathbf{v}_k.\mathbf{p}_q)^2}.
\eearr Here $\mathbf{p}_q = \hbar \mathbf{q}$ and $\mathbf{v}_k$
is the velocity of the atoms with momentum $\mathbf{k}$, $f(q) =
\sin^{-1}(\beta)/[\beta(1-\beta^2)^{1/2}]$ and
 $\beta^2 = [\omega^2 -
(\mathbf{v}_k.\mathbf{p}_q)^2]/(4\Delta^2)$.  The symbol $\langle
X \rangle$ implies averaging  of a function $X$ over the chemical
potential surface:  $ \langle X \rangle = [N(0)]^{-1}\int
d^3\mathbf{k}\delta(\epsilon_k)  X$. Note that the Eq.
(\ref{chinew}) applies to the single-particle excitation regime
($\beta^2>1$) only.
\begin{figure}
\includegraphics{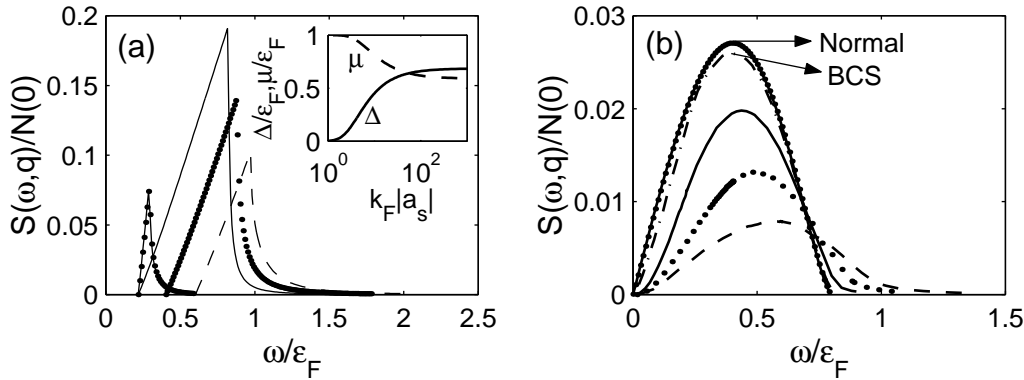}
 \caption{(a) Dimensionless DSF $S(\omega,{\mathbf
 q})/N(0)$ for single-particle excitations of a uniform superfluid Fermi gas
 is plotted as a function of dimensionless energy
 transfer $\omega/\epsilon_F$
 for different values of the scattering length
 $|a_s| = 2.76 k_F^{-1}$ (solid),  $|a_s| = 3.89 k_F^{-1}$ (dotted), $|a_s| = 5.47 k_F^{-1}$ (dashed)
 for a fixed momentum transfer $q = 0.8 k_F$. The dash-dotted
 curve is plotted for $|a_s| = 2.76k_F^{-1}$ and $q = 0.4 k_F$. The inset to Fig. (a) shows the
 variation of the gap $\Delta$ and the chemical potential $\mu$ as a function
 of $|a_s|$. For large $a_s$, $\mu$ and $\Delta$ saturate at $0.59\epsilon_F$ and
 $0.68\epsilon_F$, respectively. (b) Same as in Fig. (a) but for a trapped  superfluid Fermi gas
 for a fixed momentum transfer $q = 0.8 k_F$ ($k_F$ refers to the Fermi momentum at the trap center).
  Also shown are the DSF for small $\Delta = 0.05$ (BCS) and
 $\Delta = 0$ (normal). }
 \label{fig1}
 \end{figure}

\section{Results and discussions}
Figure 3  shows $S(\omega,{\mathbf q})$   as a function of
$\omega$ for a uniform and trapped gas for different values of
$a_s$. In the case of trapped gas, we use LDA with local chemical
potential $\mu ({\mathbf r})$ determined from equation of state
of interacting Fermi atoms in a harmonic trap. When $a_s$ is
large, the behavior of $S(\delta,{\mathbf q})$ is quite different
from that of normal as well as weak-coupling BCS superfluid. This
can be attributed to the occurrence of large gap for large
$a_s$.  In contrast to the case of a uniform superfluid,
$S(\delta,{\mathbf q})$ for a superfluid trapped Fermi gas has a
structure below $2\Delta(0)$, where $\Delta(0)$ is the gap at the
trap center. As the energy transfer decreases below $2\Delta(0)$,
the slope of $S(\delta,{\mathbf q})$ gradually reduces.
Particularly distinguishing feature of $S(\delta,{\mathbf q})$ of
a superfluid compared to normal fluid is gradual  shift of the
peak as $a_s$ or $\Delta$ increases. The quasiparticle
excitations occur only when  $2\Delta({\mathbf x}) < \omega$. This
implies that, when $\omega$ is less than $2\Delta(0)$, the atoms
at the central region of the trap can not contribute to
quasiparticle response.

\section{Conclusion}

In conclusion, we have studied long-ranged pair-correlation
inherent in both the BEC and BCS states.  We have also
investigated polarization-selective light scattering in
Cooper-paired Fermi atoms as a means of estimating the gap energy.
Our results suggest that it is possible to detect the pairing gap
 by large-angle (i.e., large $q$) Bragg scattering. Small angle
polarization-selective stimulated light scattering may be useful
in exciting BA mode. The pair-correlation which may be a generic
feature of all macroscopic quantum systems with long-range order
may serve as a potential resource for may-particle robust
entanglement that is central to quantum information science.

\end{document}